\documentclass[aps,prl,twocolumn,reprint,superscriptaddress,
]{revtex4-1} 
\usepackage{graphicx}
\usepackage{color}
\usepackage{multirow}

\begin{document}
\title{Post-aragonite phases of CaCO$_{3}$ at lower mantle pressures}
\author{Dean Smith}
 \affiliation{Department of Physics and Astronomy and HiPSEC, University of Nevada Las Vegas, Las Vegas, Nevada 89154, USA}
\author{Keith V. Lawler}
 \affiliation{Department of Chemistry and Biochemistry and HiPSEC, University of Nevada Las Vegas, Las Vegas, Nevada 89154, USA}
\author{Miguel Martinez-Canales}
 \affiliation{Scottish Universities Physics Alliance (SUPA), School of Physics and Astronomy and Centre for Science at Extreme Conditions, University of Edinburgh, Edinburgh EH9 3FD, United Kingdom}
\author{Austin W. Daykin}
 \affiliation{Department of Physics and Astronomy and HiPSEC, University of Nevada Las Vegas, Las Vegas, Nevada 89154, USA}
\author{Zachary Fussell}
 \affiliation{Department of Physics and Astronomy and HiPSEC, University of Nevada Las Vegas, Las Vegas, Nevada 89154, USA}
\author{G. Alexander Smith}
 \affiliation{Department of Chemistry and Biochemistry and HiPSEC, University of Nevada Las Vegas, Las Vegas, Nevada 89154, USA}
\author{Christian Childs}
 \affiliation{Department of Physics and Astronomy and HiPSEC, University of Nevada Las Vegas, Las Vegas, Nevada 89154, USA}
\author{Jesse S. Smith}
 \affiliation{High Pressure Collaborative Access Team, Geophysical Laboratory, Carnegie Institution of Washington, Argonne, Illinois 60439, USA}
\author{Chris J. Pickard}
 \affiliation{Department of Materials Science and Metallurgy, University of Cambridge, Cambridge CB3 0FS, United Kingdom}
 \affiliation{Advanced Institute for Materials Research, Tohoku University, Sendai 980-8577, Japan}
\author{Ashkan Salamat}
 \email[Author to whom correspondence should be addressed: ]{salamat@physics.unlv.edu}
  \affiliation{Department of Physics and Astronomy and HiPSEC, University of Nevada Las Vegas, Las Vegas, Nevada 89154, USA}

\date{\today{}}

\begin{abstract}
The stability, structure and properties of carbonate minerals at lower mantle conditions has significant impact on our understanding of the global carbon cycle and the composition of the interior of the Earth. In recent years, there has been significant interest in the behavior of carbonates at lower mantle conditions, specifically in their carbon hybridization, which has relevance for the storage of carbon within the deep mantle. Using high-pressure synchrotron X-ray diffraction in a diamond anvil cell coupled with direct laser heating of CaCO$_{3}$ using a CO$_{2}$ laser, we identify a crystalline phase of the material above 40 GPa $-$ corresponding to a lower mantle depth of around 1,000 km $-$ which has first been predicted by \textit{ab initio} structure predictions. The observed $sp^{2}$ carbon hybridized species at 40 GPa is monoclinic with $P2_{1}/c$ symmetry and is stable up to 50 GPa, above which it transforms into a structure which cannot be indexed by existing known phases. A combination of \textit{ab initio} random structure search (AIRSS) and quasi-harmonic approximation (QHA) calculations are used to re-explore the relative phase stabilities of the rich phase diagram of CaCO$_{3}$. Nudged elastic band (NEB) calculations are used to investigate the reaction mechanisms between relevant crystal phases of CaCO$_{3}$ and we postulate that the mineral is capable of undergoing $sp^{2}$-$sp^{3}$ hybridization change purely in the $P2_{1}/c$ structure $-$ forgoing the accepted post-aragonite $Pmmn$ structure.
\end{abstract}


\maketitle

\section{Introduction}

Carbonates play a significant role in the global carbon cycle through the subduction of carbonate-containing oceanic slab.\cite{Thomson_2016, Huang_2011} Models dating back to the 1980s\cite{Marty_1987} notion that the majority of the Earth's carbon is stored within the planet interior,\cite{Dasgupta_2010} either in some reduced form such as diamond or graphite, or as carbides at lower mantle and core conditions.\cite{Gao_2008, Fiquet_2009} Meanwhile, the existence of carbonate inclusions in "deep" diamonds suggest stability of the minerals under mantle pressures and temperatures.\cite{Brenker_2007, Kaminsky_2013} The stability and structures of carbonates at mantle conditions is thus important in order to further our understanding of numerous geological processes. While experiments designed to investigate such processes are often extremely challenging, the development of powerful evolutionary algorithms (for example USPEX \cite{Glass_2006}) and \textit{ab initio} random structure searching (AIRSS \cite{Pickard_2006, Pickard_2011}) approaches allows for deeper insight into the structures available to these minerals at mantle conditions.

CaCO$_{3}$ transforms from its ambient pressure calcite-I ($R\overline{3}c$) to aragonite ($Pnma$) at comparatively low pressures of below 1 GPa,\cite{Biellmann_1993} and exhibits a large variety of calcite and calcite-like phases in the sub-10 GPa region \cite{Merrill_1975, Smyth_1997, Ishizawa_2013} $-$ some of which exhibit interesting chemistry at further compression.\cite{Catalli_2005} The existence of phases beyond aragonite has been postulated since dynamic compression of aragonite CaCO$_{3}$ revealed discontinuities in its shock Hugoniot at modest pressures.\cite{Vizgirda_1982} However, static compression did not reveal a post-aragonite transition until more recently, when Santillan \& Williams observed evidence of a new phase close to 50 GPa,\cite{Santillan_2004} which was attributed to an analogue of the trigonal post-witherite BaCO$_{3}$.\cite{Holl_2000} Further study by Ono \textit{et al.} dismisses the trigonal structure in favor of an orthorhombic one with space group $P2_{1}2_{1}2$.\cite{Ono_2005} Assignment of an orthorhombic post-aragonite phase was later supported by USPEX structural simulations, which suggested the $Pmmn$ supergroup, and also predicted a further transformation into a \textit{post}-post-aragonite $C222_{1}$ phase at megabar pressures.\cite{Oganov_2006} Arapan \textit{et al.} performed density functional theory (DFT) calculations on CaCO$_{3}$ and found good agreement with the $Pmmn$ and $C222_{1}$ transition pressures predicted by Ref. \onlinecite{Oganov_2006}.\cite{Arapan_2007, Arapan_2010} Evidence for \textit{post}-post-aragonite had been observed experimentally and attributed to the pyroxene-type $C222_{1}$ structure,\cite{Ono_2007}. However, further structure searches using AIRSS predicts a CaCO$_{3}$ structure in the megabar regime with a difference $-$ a $P2_{1}/c$ unit cell with pyroxene chains stacked out-of-phase, in contrast to the parallel chains in the $C222_{1}$ structure.\cite{Pickard_2015} This difference reduces enthalpy significantly \cite{Pickard_2015, Yao_2017} and is accompanied by a marked difference in Raman signature, which was used recently by Lobanov \textit{et al.} to confirm $P2_{1}/c$-h (here, the label -h originates from Ref. \cite{Pickard_2015} and refers to the higher pressure of the two predicted $P2_{1}/c$ phases with $sp^{3}$ hybridization) as the stable structure for deep mantle CaCO$_{3}$ alongside X-ray diffraction.\cite{Lobanov_2017} Additionally, the AIRSS approach predicts a second monoclinic polymorph of CaCO$_{3}$ which is stable at pressures equivalent to a depth of 1,000 km in the mantle, the $P2_{1}/c$-l structure (the label l, as originally assigned by Ref \cite{Pickard_2015}, refers to the first of the predicted sp$^2$ hybridized $P2_{1}/c$ phases), which has remained unseen in compression experiments.

\begin{table}[t!]
\caption{\label{reviewtable}Summary of experimentally-observed high-pressure phases of CaCO$_{3}$.}
\begin{ruledtabular}
\begin{tabular}{l l c c}
Phase 					& Space group 				& $P$ (GPa) 	& 		 					\\
\hline \noalign{\smallskip{}}
Calcite-I 					& $R\overline{3}c$ (no. 167)	& -- 			& 							\\
Aragonite 					& $Pnma$ 	(no. 62)			& 0.67 		& \cite{Biellmann_1993} 			\\
CaCO$_{3}$(II) 				& $P2_{1}/c$ (no. 14)		& 1.5 		& \cite{Merrill_1975} 				\\
\noalign{\bigskip{}}
\multirow{2}{*}{
CaCO$_{3}$(III)} 			& $C2$ (no. 5)				& 4.1 		& \cite{Smyth_1997} 				\\
 						& $P\overline{1}$ (no. 2)		& 2.5 		& \cite{Merlini_2012} 				\\
\noalign{\bigskip{}}
CaCO$_{3}$(IIIb) 			& $P\overline{1}$ (no. 2)		& 2.5 		& \cite{Merlini_2012} 				\\
CaCO$_{3}$(VI) 			& $P\overline{1}$ (no. 2)		& 15 			& \cite{Merlini_2012} 				\\
$P2_{1}/c$-l 				& $P2_{1}/c$ (no. 14)		& 41.3		& This study					\\
Post-aragonite 				& $Pmmn$ (no. 59)			& $\sim{}$40	& \cite{Ono_2005, Oganov_2006} 	\\
$sp^{3}$-CaCO$_{3}$ 		& $P2_{1}/c$ (no. 14)		& 105 		& \cite{Lobanov_2017} 			\\
\end{tabular}
\end{ruledtabular}
\end{table}

In a single-crystal X-ray diffraction study, Merlini \textit{et al.} detect a triclinic $P\overline{1}$ structure (CaCO$_{3}$-VI) above 15 GPa \cite{Merlini_2012} which they attribute to the same transition detected, but not indexed, by previous dynamic compression experiments.\cite{Vizgirda_1982} Interestingly, this triclinic phase had previously been predicted by the UPSEX code,\cite{Oganov_2006} but dismissed by those authors as metastable with respect to $Pnma$ aragonite. This finding was echoed in Ref. \onlinecite{Pickard_2015}, which found CaCO$_{3}$-VI to be higher in enthalpy than aragonite and intrinsically strained, in spite of Ref. \onlinecite{Merlini_2012} measuring a higher density for CaCO$_{3}$-VI than aragonite during their experiments. However, that CaCO$_{3}$ is able to occupy numerous metastable and transient phases at modest\cite{Merrill_1975, Smyth_1997, Catalli_2005, Ishizawa_2013} and high\cite{Merlini_2012, Pickard_2015} pressures  $-$ much like its rich ground-state phase progression under compression $-$ is a testament to its remarkably diverse polymorphism. We can attribute the richness of the CaCO$_{3}$ phase diagram to the close matching in size of the Ca$^{2+}$ and CO$_{3}^{2-}$ species. Structural ramifications of the ionic size ratio in $M$CO$_{3}$ aragonite group crystals ($M$ = Ca, Sr, Ba, Pb) is exemplified by observations made by Ref. \onlinecite{Antao_2009}, where combined high-resolution X-ray diffraction and neutron diffraction record an increasing degree of disorder in CO$_{3}^{2-}$ units with decreasing cation size, with CaCO$_{3}$ having the largest variation in C-O bond lengths as well as a deviation from truly planar carbonate ions as a result of steric effects. Indeed, in the case of MgCO$_{3}$, this ionic size effect inhibits the formation of an aragonite phase anywhere in its phase diagram and stabilises the $R\overline{3}c$ structure up to 85 GPa.\cite{Pickard_2015} Similarly, for heavier carbonates where the sites occupied by CO$_{3}^{2-}$ grow with $M$, we expect a comparatively simple phase evolution with pressure.

That the phase diagram of such a common and important mineral $-$ and one exhibiting a wide array of stable structures which are relevant to geological processes in the Earth's mantle $-$ has only begun to be unraveled experimentally since the turn of the century\cite{Santillan_2004, Ono_2005, Ono_2007, Lobanov_2017} is largely a result of experimental advances allowing powerful diagnostics such as X-ray diffraction to be performed \textit{in situ} at combined high pressure and temperature.\cite{Prakapenka_2008, Salamat_2014, Petitgirard_2014, Meng_2015} A summary of the experimentally-confirmed high-pressure phases of CaCO$_{3}$ is given in Table \ref{reviewtable} alongside their onset pressures. 
 Here, we present the addition of a transitional structure experimentally realized in CaCO$_{3}$ by utilizing a recently-developed instrument to allow \textit{in situ} CO$_{2}$ laser annealing of minerals at high pressure $-$ the $P2_{1}/c$-l phase previously predicted by AIRSS calculations,\cite{Pickard_2015} which exists as an intermediate between aragonite and ``post-aragonite''.

\section{Structure predictions}

\begin{figure*}
\includegraphics[width=17.8 cm]{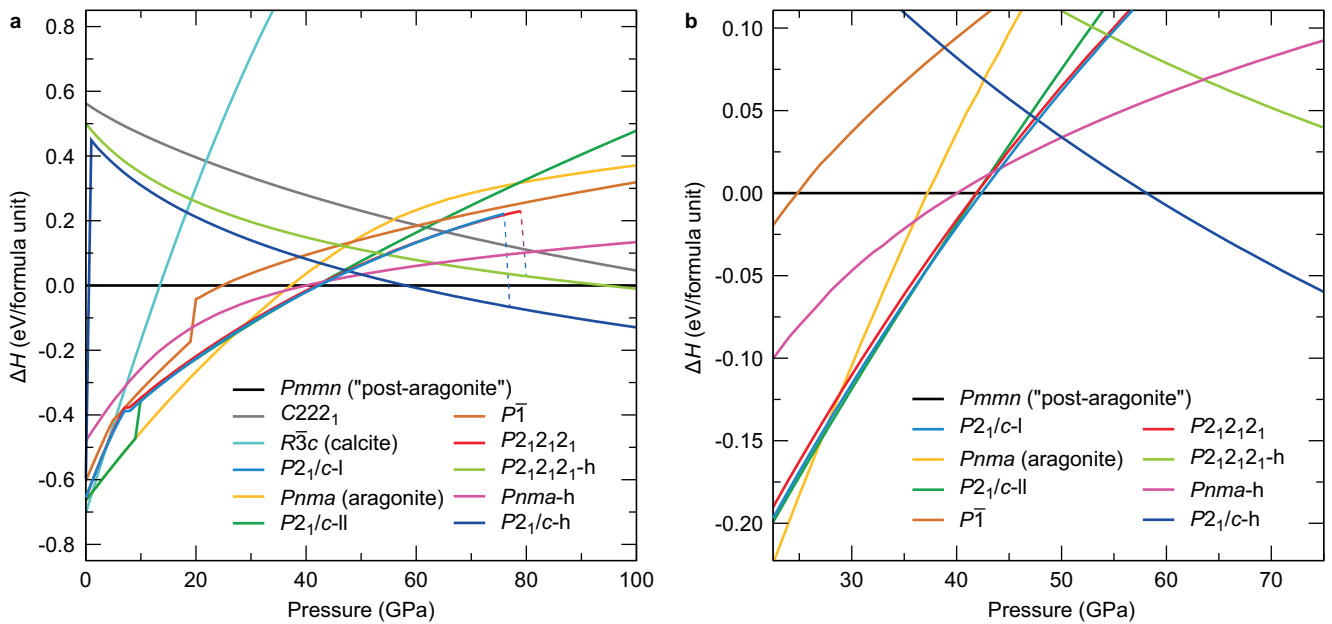}
\caption{\label{PH} Enthalpy per formula unit of CaCO$_{3}$ phases relative to ``post-aragonite'' $Pmmn$, $\Delta{}H$, as a function of pressure (a) to 100 GPa, and (b) around the transitional post-aragonite region. Dashed lines indicate structures which collapse into high pressure phases. Pale lines indicate structures which are never thermodynamically stable.}
\end{figure*}

We first revisit the \textit{ab initio} random structure search (AIRSS) for CaCO$_{3}$ to 100 GPa in Ref. \onlinecite{Pickard_2015} by re-evaluating the initially reported structures and performing subsequent searches to uncover more candidate structures. Shown in Fig. \ref{PH}, the enthalpy of the structures was computed by fully relaxing over a range of pressures with the Perdew-Burke-Ernzerhof for solids and surfaces (PBEsol) generalized gradient approximation (GGA) \cite{Perdew_2008} density functional using the CASTEP \cite{ClarkStewart_2009} plane-wave DFT code. The choice of PBEsol as the density functional is taken as the calculations predict a calcite-I $\rightarrow{}$ aragonite transition pressure of 1.4 GPa, which is closer to experimental observations (0.67 GPa \cite{Biellmann_1993}) than the 4 GPa prediction attained when the Perdew-Burke-Ernzerhof (PBE) functional was used. The PBEsol functional, in general, provides better accuracies for predicted volumes of condensed systems. 
The basis set cutoff energy was set to 700 eV using ultrasoft pseudopotentials with valence configurations of $3s^{2}3p^{6}4s^{2}$ for Ca, $2s^{2}2p^{2}$ for C, and $2s^{2}2p^{4}$ for O.\cite{Vanderbilt_1990} A Monkhorst-Pack \cite{Monkhorst_1976} $k$-point grid with spacing 0.3 2$\pi$ \AA{}$^{-1}$ was used to sample the Brillouin zone. The results presented in Fig. \ref{PH} are in very good agreement with other DFT results.\cite{Arapan_2007, Arapan_2010, Yao_2017} 

Beyond the stability field of $Pnma$ aragonite, we observe two competing monoclinic structures $-$ the $P2_{1}/c$-l structure from Ref. \cite{Pickard_2015}. which collapses into $P2_{1}/c$-h at high pressures (blue dashed line in Fig. \ref{PH}), as well as a second which we name $P2_{1}/c$-ll. Interestingly, the $P2_{1}/c$-ll phase which has comparably low enthalpy at low pressures to aragonite is not the previously known CaCO$_{3}$-II $P2_{1}/c$ structure.\cite{Merrill_1975} $P2_{1}/c$-l and $P2_{1}/c$-ll are enthalpically competitive above 10 GPa, crossing one another in stability once at 18 GPa and once again at 37.5 GPa, whereupon the $P2_{1}/c$-l remains the most stable until its collapse into $P2_{1}/c$-h. The maximum separation of the two competing structures is 3.23 meV/formula unit (f.u.) (0.074 kcal mol$^{-1}$ f.u.$^{-1}$). Both the $P2_{1}/c$-l and $P2_{1}/c$-ll phases become more enthalpically more favorable than aragonite above 27.2 GPa. A further competitive phase is found in this region with $P2_{1}2_{1}2_{1}$ symmetry, and which eventually collapses into $sp^{3}$-bonded $P2_{1}2_{1}2_{1}$-h, however neither of these structures occupy the lowest enthalpy at any pressure. The previously reported $Pnma$-h once again appears as a competitive phase, and is stable relative to aragonite and each of the aforementioned phases above 44 GPa, but at this point is not stable relative to $Pmmn$. In ARISS, we observe a more narrow field of stability for $Pmmn$ ``post-aragonite'' than our past study (42.4$-$58 here compared with 42$-$76 GPa in Ref. \cite{Pickard_2015}) before $P2_{1}/c$-h becomes the dominant phase until at least 100 GPa, as has recently been confirmed by experiments performed by Lobanov \textit{et al.},\cite{Lobanov_2017} however the AIRSS method does not account for temperature effects $-$ in later sections we have explored the relative stabilites in more depth using quasi-harmonic approximation (QHA) calculations.

\begin{figure}
\includegraphics[width=\columnwidth]{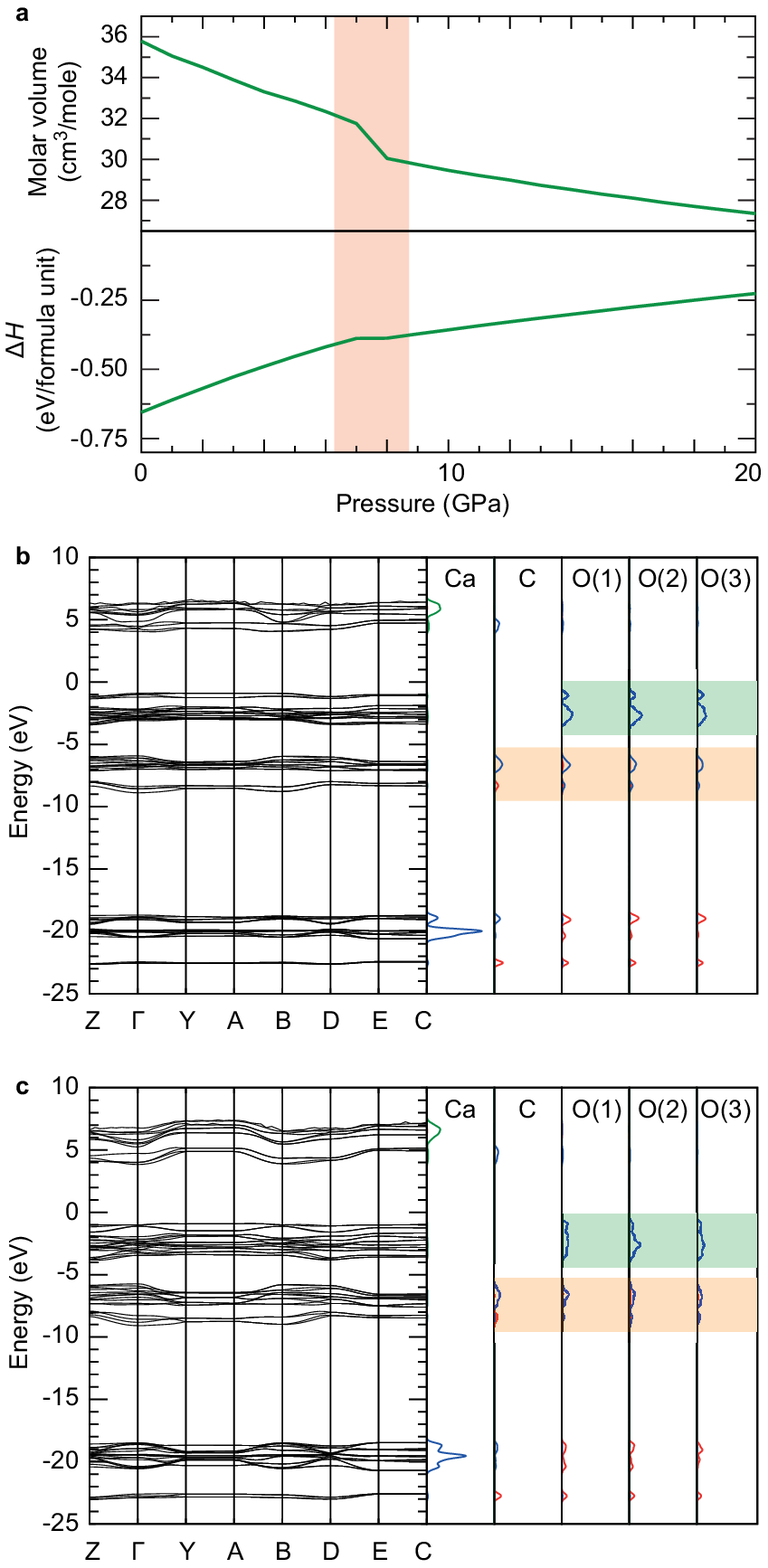}
\caption{\label{bands} Electronic evolution of $P2_{1}/c$-l CaCO$_{3}$ at 20 GPa. (a) Relative enthalpy, $\Delta{}H$, and volume display discontinuous behavior (relative to Pmmn as in Fig. \ref{PH}), (b,c) Calculated band structure (left) and partial density of states (right) at 0 and 10 GPa (pDOS due to $s$, $p$ and $d$ electrons are shown in red, blue and green respectively). The changes in the electronic structure as a function of pressure may be observed for the oxygen lone pair states (highlighted in green) and the C-O sp$^2$ bonding states (highlighted in orange) and their corresponding bands to the immediate left.}

\end{figure}

DFT computations for $P2_{1}/c$-l CaCO$_{3}$ reveal a discontinuity in its enthalpy with pressure at around 7.5 GPa, which is reflected in the pressure-volume equation of state of the phase (Fig. \ref{bands}a). As such behavior can be related to transitions, we investigate the structure and electronic structure of the $P2_{1}/c$-l phase at low pressures. Fig. \ref{bands} shows the band structure and partial density of states (pDOS) of CaCO$_{3}$ in the $P2_{1}/c$-l phase at ambient pressure and 10 GPa (Fig. \ref{bands}b and c respectively). At ambient conditions, we see the signature of a resonant $sp^{2}$ CO$_{3}^{2-}$ structure, $s$ and $p$ lobes which are near-symmetric (orange bands in Fig. \ref{bands}b and c) and a valence composed of the remaining oxygen lone pairs tightly localized in energy close to the Fermi level (green bands). When pressure is increased to 10 GPa, the $sp^{2}$ features become smeared and more asymmetric, and are accompanied by additional curvature in the bands indicative of a changing bonding environment. Physically, this is accompanied by a puckering of CO$_{3}^{2-}$ groups. The same is true for the oxygen lone pair features at the top of the valence band due to its splitting, as well as for the Ca $p$ feature around -20 eV. These behaviors are suggestive of an interaction (such as polarization or charge transfer) at higher pressures between the Ca$^{2+}$ and CO$_{3}^{2-}$ species, and is supported further by the enhanced curvature of the conductance band at 10 GPa.

\section{High pressure experiments}

High pressure experiments were performed in diamond anvil cells (DACs) of custom design, equipped with conical-cut diamonds with a 70$^{\circ}$ opening and 300, 200 and 100 $\mu$m culets for three separate runs. Re foil with an initial thickness of 200 $\mu$m was pre-indented to form a gasket, and a 180, 120 and 60 $\mu$m hole were drilled, respectively, to serve as the sample chamber by laser micromachining.\cite{Hrubiak_2015} CaCO$_{3}$ powder (Sigma-Aldrich ReagentPlus $\geq$ 99\%) was oven-dried and pressed into 10 $\mu$m-thick pellets, with an average diameter of 40 $\mu$m. High pressure experiments typically have the requirement that samples are surrounded by some soft medium to serve as a quasi-hydrostatic pressure transmitter, and due to the high thermal conductivity of diamond, laser-heated DAC experiments require that samples are thermally isolated from the diamond anvils to achieve efficient and homogeneous heating.\cite{Petitgirard_2014} Thus, CaCO$_{3}$ pellets were encased in either a medium of NaCl, KBr  or Ar respectively, whose well-calibrated equation of state is also used to calculate pressure inside the sample chamber.\cite{Dorogokupets_2007, Dewaele_2012, Ross_1986}

\begin{figure}
\includegraphics[width=\columnwidth]{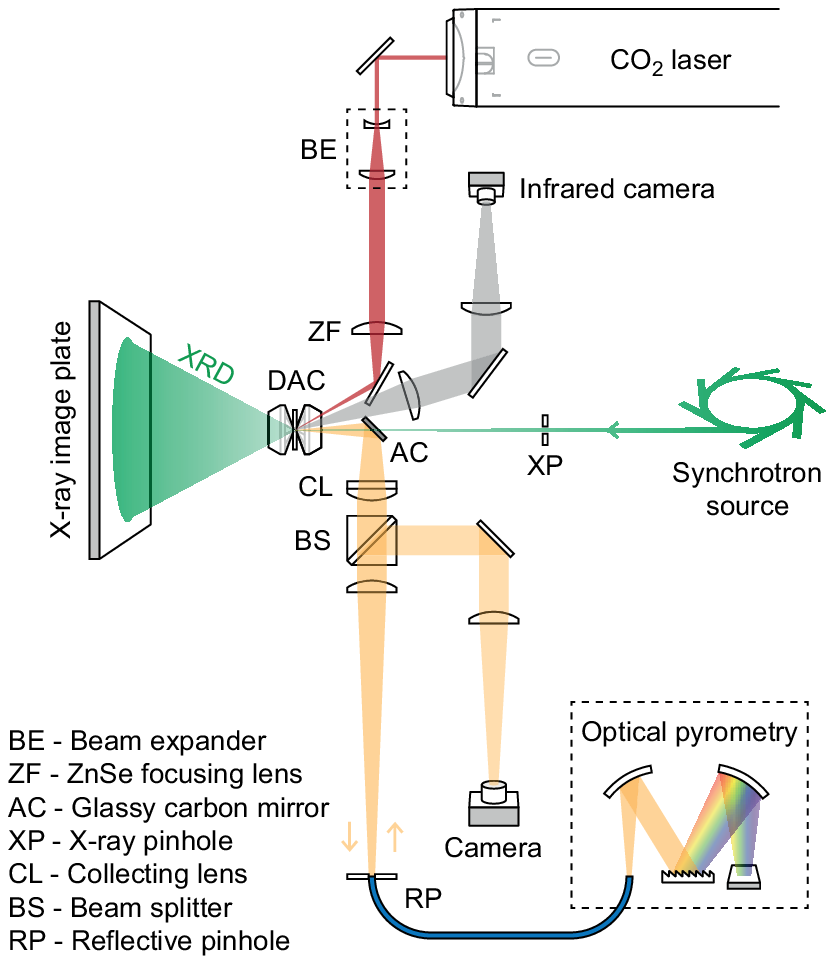}
\caption{\label{co2lh} \textit{In situ} CO$_{2}$ laser heating system at the HPCAT ID-B diffraction beamline. Optical paths are shown for CO$_{2}$ laser delivery in red, pyrometry and visible spectrum visualization in yellow, mid-IR visualization in grey, and synchrotron X-ray diffraction in green.}
\end{figure}

Laser heating was performed \textit{in situ} at the ID-B diffraction beamline at HPCAT (Sector 16, Advanced Photon Source, Argonne IL) using the recently-developed instrument depicted in Figure \ref{co2lh}. A Synrad \textit{evolution}125 CO$_{2}$ laser operating at 10.6 $\mu$m was focused into the DAC to a spot size 35-80 $\mu$m. Thermal emission is collected using an achromatic objective shielded by a MgF$_{2}$ window to protect the glass optics from damage by the diffuse 10.6 $\mu$m radiation, and the image is then refocused onto a 50 $\mu$m pinhole made in a reflective surface. This pinhole acts as a spatial filter which samples a 7.5 $\mu$m region of the sample space, comparable to the X-ray spot size, and is aligned to the peak in the X-ray fluorescence such that temperature measurements are made from the precise location of X-ray diffraction. Reflected light from the pinhole surface is imaged onto a CCD camera for viewing the DAC interior during experiments.

For accurate alignment of the mid-IR laser spot to the X-ray and pyrometer focus, we employ a thermal imaging camera modified for microscopy (grey paths in Fig. \ref{co2lh}). In this way, we are able to directly image the sample chamber in the 7$-$14 $\mu$m region prior to heating. This crucial development allows alignment of the laser spot to the sample area within the DAC using only milliwatts of laser power, whereas previous CO$_{2}$ laser heating instruments for DAC experiments have relied heavily on laser radiation coupling with material inside the sample chamber to create a thermal glow to allow the laser spot to be located, pre-heating the sample environment in some cases to in excess of 1,500 K. Implementation of mid-infrared microscopy to directly visualize the 10.6 $\mu$m laser spot makes it possible to avoid any pre-heating of the sample during the alignment procedure and before the formal beginning of an experiment.\cite{HPCATCO2LH}

\begin{figure}
\includegraphics[width=\columnwidth]{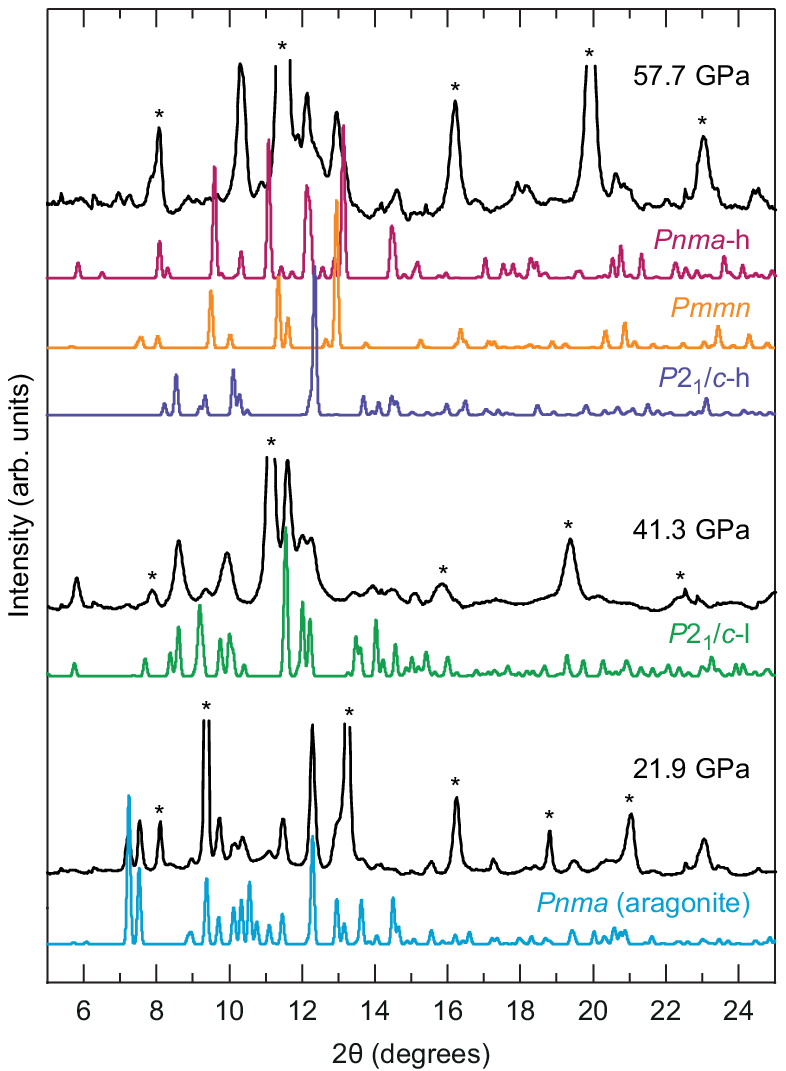}
\caption{\label{stack} Observed X-ray diffraction patterns at 21.9, 41.3 and 57.7 GPa following annealing with CO$_{2}$ laser (black plots), normalised to maximum intensity from CaCO$_{3}$. Peaks marked with asterisks (*) are due to NaCl thermal insulator and pressure medium. Simulated diffraction patterns at $\lambda$ = 0.4066 \AA{} and equivalent unit cell volumes for each phase are stacked beneath using computationally-generated structures. Annealed CaCO$_{3}$ data at 57.7 GPa is stacked against the three lowest enthalpy phases at this pressure and, noteably, there is little similarity between the observed pattern and the ``post-aragonite'' $Pmmn$ phase, nor with other competitive structures.}
\end{figure}

Angle-dispersive X-ray diffraction patterns were collected both during and after laser heating using monochromatic X-rays with $\lambda$ = 0.4066 \AA{}, with a Pilatus $1M$ detector.\cite{Broennimann_2006} The two-dimensional images were integrated into one-dimensional spectra using the \textsc{Dioptas} software package.\cite{Prescher_2015} We gradually increased power density from the CO$_{2}$ laser as X-ray diffraction patterns were collected and observe for changes in the sample structure and texturing as a function of time and laser power. Temperature determination by optical pyrometry when working with insulating materials such as carbonate minerals is complicated by their low emissivity in the visible, which remains low even into the mid-IR.\cite{Lane_1999, Glotch_2013} The pyrometer is sufficiently sensitive to detect X-ray fluorescence from the sample and salt medium during diffraction collections, but does not detect any thermal emission from the CaCO$_{3}$ even when heating is performed at the highest power densities, making temperature estimation wrought with uncertainty. In Fig. \ref{stack}, we show select data from the phase progression of CaCO$_{3}$ to 57.7 GPa, with laser annealing performed at roughly 5 GPa steps in pressure. Below 40 GPa, we observe only the expected Bragg reflections from aragonite.

Although marginally lower in enthalpy than all competing phases between 27.2 GPa and 37.5 GPa, we did not observe the $P2_{1}/c$-ll structure when annealing at 30.6 or 36.2 GPa $-$ further inspection into free energy arguments is sought to help understand this discrepancy. At 40 GPa, $P2_{1}/c$-ll exhibits the shortest Ca-Ca distance of any of the studied structures $-$ 3.17 \AA{} \textit{vs.} 3.47 in $Pnma$ aragonite and 3.65 in $P2_{1}/c$-l. This is a markedly compressed distance for two of the heaviest atoms in the cell, and likely translates to a higher vibrational potential energy of any of the observed structures from a harmonic oscillator point-of-view. In order to account for this, as well as the effects of low temperature on the relative stability of phases, we have computed the Gibbs free energies of the most competitive structures within the quasi-harmonic approximation (QHA).

\begin{figure}
\includegraphics[width=\columnwidth]{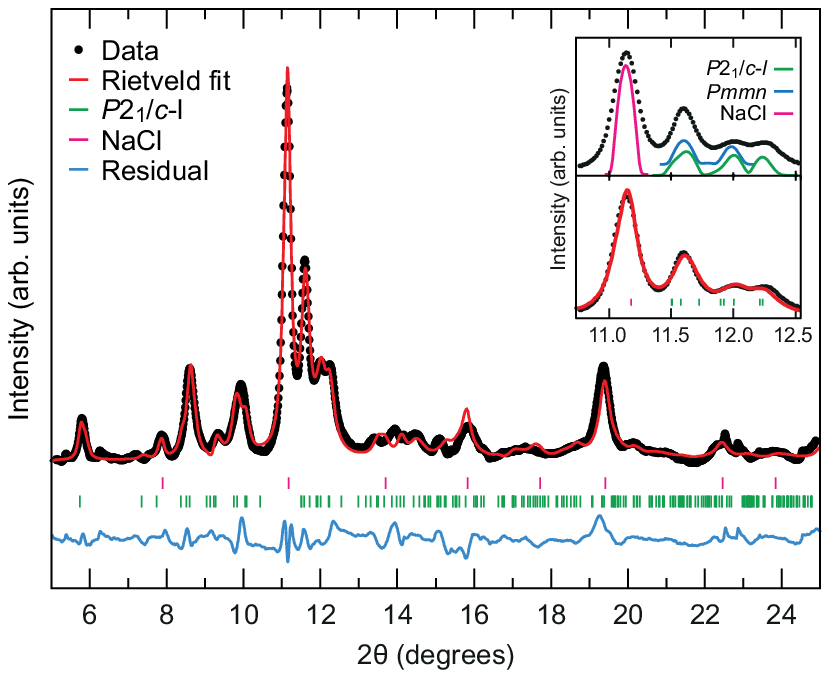}
\caption{\label{P21c} Results of Rietveld structural refinement of X-ray diffraction pattern of CaCO$_{3}$ at (a) 41.3 GPa. A good fit ($wRp$ = 1.06\%) is achieved using the $P2_{1}/c$-l phase predicted by Ref. \onlinecite{Pickard_2015}. (inset) Simulated X-ray diffraction patterns for $P2_{1}/c$-l and $Pmmn$ structures compared with observed data (top) and Rietveld fit (bottom) to peaks in the 12$^{\circ}$ $2\theta$ region using the $P2_{1}/c$-l structure.}
\end{figure}

At 41.3 GPa, a change in the diffraction pattern is observed, and the resulting pattern cannot be accounted for with the $Pnma$ aragonite or $Pmmn$ post-aragonite structures. We index the diffraction pattern at 41.3 GPa as corresponding to the monoclinic $P2_{1}/c$-l structure, found by DFT calculations to be the most stable structure at 37.5$-$42.4 GPa (at 0 Kelvin). Fig. \ref{P21c} shows the results of a Rietveld refinement of the crystal structure at 41.3 GPa using the GSAS software package with the $P2_{1}/c$-l structure type as well as the high-pressure $Pm\overline{3}m$ structure of the NaCl thermal insulator. The full structural refinement on CaCO$_{3}$ has weighted $R$-factor value of 1.06\% and a reduced $\chi^2$ value of 4.202. This crystal structure exhibits a strong peak at around 5.75$^{\circ}$ $2\theta$ due to Bragg reflections from the (100) planes, which disappears beyond 50 GPa signifying a further transition. The inset of Fig. \ref{P21c} shows the group of observed Bragg reflections around 12$^\circ$ $2\theta$ alongside simulated patterns for the $P2_{1}/c$-l and $Pmmn$ phases (top panel), which show distinct differences in this region. The $Pmmn$ ``post-aragonite'' structure (red) predicts two features between 11.5-12.0$^{\circ}$, whereas the observed diffraction pattern contains three, which the $P2_{1}/c$-l structure (green) is able to account for $-$ further evidenced by the ability of this phase to accurately model the data in this region during Rietveld refinement (bottom panel). Table \ref{table} shows our recorded experimental $P2_{1}/c$-l structure from Rietveld refinement in which only the positions of the Ca atoms were refined and the $P2_{1}/c$-l structure from simulation at equivalent pressures.

Above 50 GPa, there is a transition away from $P2_{1}/c$-l. However, as is evident from the observed diffraction pattern in Fig. \ref{stack} and the simulated $Pmmn$ diffraction pattern using appropriate lattice parameters (orange line in Fig. \ref{stack}), the pattern above 50 GPa cannot be indexed with the ``post-aragonite'' phase. This was sustained until 70 GPa with laser annealing performed at $\sim{}$5 GPa intervals, with no evidence for improvements in crystallinity nor further structural transitions. Subsequent decompression $-$ again with frequent laser annealing $-$ saw the system readopt the $Pnma$ aragonite structure once below 35 GPa.

Upon reflection in the literature, the accepted $Pmmn$ structure for post-aragonite CaCO$_{3}$ still remains to be subject to a rigorous structural refinement $-$ both the original assignment of orthorhombic $P2_{1}2_{1}2$ by Ref. \onlinecite{Ono_2005} and the following reassignment to $Pmmn$ by Ref. \onlinecite{Oganov_2006} feature the same raw data and compare it with simulated patterns without performing any fitting procedure. Ref. \onlinecite{Oganov_2006} provides atomic positions for their experimental pattern, but there is no evidence for a structural refinement having been performed on the data. Until very recently, the raw data in Refs. \onlinecite{Ono_2005} and \onlinecite{Oganov_2006} was the only published evidence of X-ray diffraction from $Pmmn$ CaCO$_{3}$, in spite of its recurrence in the literature.\cite{Ono_2007, Arapan_2007, Arapan_2010} Lobanov \textit{et al.} have since heated CaCO$_{3}$ at 83 GPa using indirect laser-heating methods and published the raw diffraction pattern against simulated peaks. That the $Pmmn$ phase is clearly not observed in our study warrants further investigation, and will be discussed in later sections.

During the revisions to this manuscript, a study on post-aragonite CaCO$_{3}$ was published in Ref. \onlinecite{Gavryushkin_2017} which features indirect laser heating of CaCO$_{3}$ in a similar pressure regime. Gavryushkin \textit{et al.} present the discovery of a new phase, termed ``aragonite-II'', at 35 GPa and present ``unambiguous'' evidence for the $Pmmn$ structure of post-aragonite at 50 GPa. It is important to consider the quality of the data behind these claims. In both cases, the phases present are subject to a Pawley (\textit{i.e.} structureless) refinement over a very small range of reflections. Ref. \onlinecite{Gavryushkin_2017} present their new ``aragonite-II'' phase alongside a further proposed new structure, termed CaCO$_{3}$-VII: a portmanteau of the $P2_{1}/c$-h unit cell and $P2_{1}/c$-l atomic positions (though, more accurately, the $P2_{1}/c-h$ unit cell, since the Pawley refinement employed is insensitive to atomic positions). The CaCO$_{3}$-VII structure conveniently fits a number of the reflections which are not accounted for by the aragonite-II unit cell. The same CaCO$_{3}$-VII structure persists at 50 GPa alongside the $Pmmn$ post-aragonite, where it accounts for only a single prominent reflection. The observations in Ref. \onlinecite{Gavryushkin_2017} of mixed-phase CaCO$_{3}$ at 50 GPa reinforces our discussion in later sections regarding the kinetics of $Pmmn$ post-aragonite.

It should also be noted that the metastable CaCO$_{3}$-VI phase observed in Ref. \cite{Merlini_2012} was not observed at any point during our high pressure experiment, in spite of their finding that it is higher in density than aragonite up to 40 GPa. The observation of metastable CaCO$_{3}$ phases by compression is not uncommon, very early diamond anvil cell high pressure studies uncovered the structure of CaCO$_{3}$-II at 1.5 GPa.\cite{Merrill_1975} Such metastable phases are likely avoided using our CO$_{2}$ laser annealing approach, since direct heating with 10.6 $\mu$m radiation allows uniform heating and homogeneous phase transformations at each density, allowing kinetic barriers that may lead to sluggish phase transitions and the development of metastable structures to be overcome with a high degree of control. Compared with alternative laser-heating methods at high pressure, which employ a metallic coupling material to strongly absorb $\sim{}$ 1 $\mu$m radiation and indirectly heat the sample material, the CO$_{2}$ laser heating approach is a more close analogue of the geothermal annealing experienced by real mantle minerals. Future experiments could achieve highly accurate determination of ground state structures in minerals by combining the CO$_{2}$ laser annealing method employed here with the single crystal and multi-grain methodologies employed to solve more complex polymorphs such as the metastable CaCO$_{3}$-VI,\cite{Merlini_2012} and while necessary preparations for such experiments are challenging even with near-IR laser irradiation,\cite{Merlini_2015} it is the most natural progression of high $P$,$T$ experiments on geological materials to allow for the most accurate measurements. 

\begin{figure}
\includegraphics[width=\columnwidth]{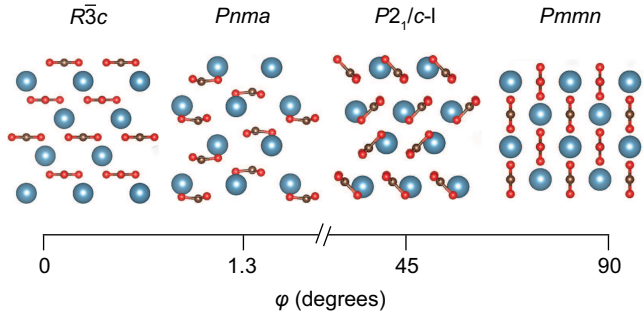}
\caption{\label{CO3angle} Angle $\varphi$ between CO$_{3}^{2-}$ groups and Ca stacking in CaCO$_{3}$ crystal structures.}
\end{figure}

In Fig. \ref{CO3angle}, we show the ambient calcite ($R\overline{3}c$) and aragonite ($Pnma$) structures of CaCO$_{3}$ alongside the $P2_{1}/c$-l phase from this study and the higher pressure ``post-aragonite'' ($Pmmn$ \cite{Ono_2005}), and follow the angle $\varphi$ between CO$_{3}^{2-}$ groups in each phase and their \textit{bc} stacking plane in $R\overline{3}c$. In the ambient calcite and aragonite phases, a distinct ordering of the CO$_{3}^{2-}$ units along these planes is evident. The CO$_3^{2-}$ groups in aragonite are likely more ordered than is shown, since the aforementioned puckering of CO$_{3}^{2-}$ units appears to exaggerate the apparent rotation with respect to the stacking plane. These co-planar groups are responsible for the anisotropic nature of a number of material properties in bulk CaCO$_{3}$ $-$ for instance, the elastic constants of calcite vary by a factor of 4.5 across its crystallographic directions and the shear wave velocity by a factor of 1.5,\cite{Chen_2001} and natural CaCO$_{3}$ crystals famously exhibit strong birefringence and polarizing properties.\cite{Ropars_2011} Co-planar CO$_{3}^{2-}$ is present in the high-pressure CaCO$_{3}$-II and CaCO$_{3}$-III phases ($C2$ \textit{c} axis \cite{Merrill_1975} and $P2_{1}/c$ \textit{b} axis \cite{Smyth_1997} respectively, not shown), and in the $Pmmn$ post-aragonite phase. In the intermediate $P2_{1}/c$-l phase, however, carbonate ions are at 45$^\circ{}$ to the $R\overline{3}c$ stacking plane and stack $AABB$ with one another. One could expect, then, that the physical properties of $P2_{1}/c$-l CaCO$_{3}$ differ from those of calcite and aragonite, having a more isotropic elastic stiffness. Such crystallographic observations in deep Earth minerals can prove useful in advising geology and seismology, where mechanical properties of mantle constituents are directly recorded by sound velocities within the planet. That the carbonate ions occupy an intermediate angle in $P2_{1}/c$-l CaCO$_{3}$ correlates with the structure occupying the energetic intermediate between aragonite and $Pmmn$, and that some intermediate featuring non-co-planar CO$_{3}^{2-}$ is not surprising since in the absence of this intermediate phase, the carbonate ions in CaCO$_{3}$ would be required to undergo a near 90$^{\circ}$ rotation during the post-aragonite phase transition.

\section{Quasi-harmonic approximation}

The contribution of temperature to the phase stabilities of CaCO$_{3}$ has been explored by computation of the Gibbs free energies of the most competitive phases between 30 and 80 GPa: $Pnma$ (aragonite), $P2_{1}/c$-l, $P2_{1}/c$-ll, $Pmmn$ (post-aragonite) and $P2_{1}/c$-h. We have followed an identical procedure as that outlined in Ref. \onlinecite{Ackland_2017}. We have computed the phonon spectra at volumes covering DFT pressures between at least 35 and 60 GPa every 5 GPa. Furthermore, spectra down to 25 GPa for the low energy phases, and up to 75 GPa for the high pressure ones have been computed. All phonon calculations have been performed with \textsc{Quantum-Espresso} 6.1,\cite{Giannozzi_2009} with ultrasoft pseudopotentials similar to the CASTEP ones, also generated with the PBEsol exchange-correlation functional. As a sanity check, the relative enthalpies are in excellent agreement with those calculated with CASTEP.

\begin{figure}
\includegraphics[width=\columnwidth]{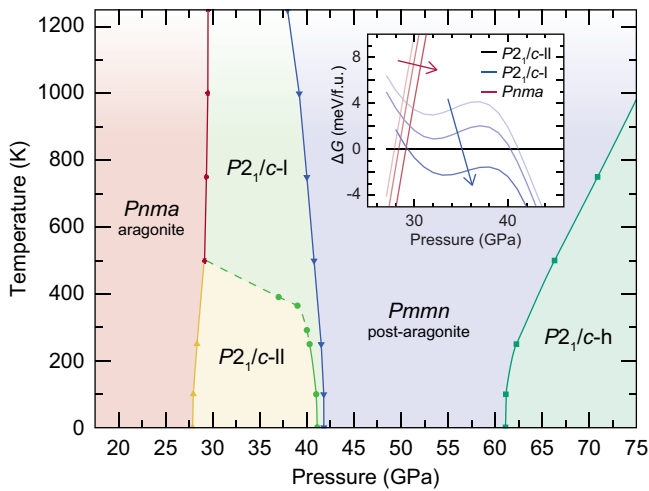}
\caption{\label{QHA} Calculated pressure-temperature phase diagram of CaCO$_{3}$, within the quasi-harmonic approximation. The main effect of temperature is stabilizing post-aragonite against all other structures, and favors $P2_{1}/c$-l over $P2_{1}/c$-ll at the temperatures over 500 K. (inset) $\Delta{}G$ for the aragonite $Pnma$ and competitive $P2_{1}/c$-l and $P2_{1}/c$-ll structures under pressure, arrows denote the temperature increase 0, 250 and 500 K.}
\end{figure}

The introduction of vibrational effects $-$ as seen in Fig. \ref{QHA} $-$ has only a small influence on relative stabilities below room temperature. Above that, however, the contribution of vibrational effects stabilizes $Pmmn$ over all other candidate structures, especially $P2_1/c$-h. However, it is worth nothing that below 28 GPa an entire branch of the phonon dispersion along the path connecting the $\Gamma{}$ and Z special points in the Brillouin zone becomes zero, suggesting a lack of dynamic stability of the structure at these pressures.

As alluded to in the previous section, temperature also stabilizes $P2_{1}/c$-l over $P2_{1}/c$-ll. The inset to Fig. \ref{QHA} shows the computed Gibbs free energy of the two competitive phases, alongside aragonite, at temperatures of 0, 250 and 500 K. At 250 K, there is a separation between $P2_{1}/c$-l and $P2_{1}/c$-ll of $\sim{}$ 2 meV/formula unit, which is small enough to be comparable with the error in the QHA calculations, and this persists until 500 K, where the $P2_{1}/c$-l phase is ultimately lower in energy $-$ the dotted green line in Fig. \ref{QHA} shows the region on the phase diagram where the $G$ of $P2_{1}/c$-l and $P2_{1}/c$-ll are within this limit. That the $P2_{1}/c$-ll phase is so stable until temperature effects are considered may suggest that aragonite may undergo a transformation into this structure at low temperatures. For comparison, it was recently demonstrated in pure lithium that, at ambient pressure, the face-centered cubic polymorph is only accessible at low temperatures through hysteretic pathways, despite it being the energetic ground state.\cite{Ackland_2017} With this in mind and with the energetic competitiveness of the $P2_{1}/c$-l and $P2_{1}/c$-ll structures at moderate temperatures, there is the potential that $P2_{1}/c$-ll exists as a low-temperature phase in CaCO$_{3}$ $-$ further implied by its absence from the literature despite several decades of room-temperature compression. Rather, we observed aragonite at 30.6 and 36.2 GPa in our annealing experiments, suggesting stability of the aragonite $Pnma$ phase to at least 36.2 GPa; however, the combined uncertainties of theory and experiment (originating from choice of pressure medium and its calibration) could amount to $\pm{}$5 GPa on the phase line connecting $Pnma$ and the new $P2_{1}/c$-l structures.

We note that the vibrational pressure $\frac{\partial F_{\textrm{vibr}}}{\partial V}$ is very substantial beyond 1250 K for the low pressure monoclinic phases, and therefore the QHA results should not be relied upon beyond that point. Against our experimental observations, the QHA predicts a wider stability range for $Pmmn$. Discarding thermal stabilization of other known competitive structures, this points at the large kinetic barrier which we have explored using NEB in the following section. A large barrier is also consistent with a significant difference in molar volumes: at 45 GPa, $Pmmn$ has a theoretical volume 3.5\%{} smaller than that of $P2_{1}/c$-l, roughly equivalent to 5 GPa.

\section{Transformation Mechanisms}

To elucidate why the $P2_{1}/c$-l phase has not previously been observed, and why it is realized only in a narrow pressure range, we carried out generalized solid-state nudged elastic band (g-SSNEB \cite{Sheppard_2012}) simulations at 20, 40, and 60 GPa along the pathway shown in Fig. \ref{CO3angle}. The reaction pathways were calculated with the Vienna \textit{ab initio} simulation package (VASP) version 5.4.1 modified for g-SSNEB. The VASP calculations were set up similarly to the CASTEP calculations with the exceptions of the use of a $\Gamma$-centered $k$-point grid and the projector augmented wave (PAW \cite{Bloechl_1994}) method to describe the electron-ion interactions. The g-SSNEB calculations connected a single unit cell of each phase co-oriented to follow the pathway described in Figure \ref{CO3angle}, and they employed 24 images to connect the initial and final phases. Both the atomic positions and lattice vectors of the images were allowed to vary. The g-SSNEB simulations ran until the forces were below 10$^{-2}$ eV/\AA{}. The AIRSS predicted structures were re-optimized in VASP prior to g-SSNEB to minimize any inconsistencies between different DFT implementations, and the defined states maintained the same energy orderings as in Fig. \ref{PH}.

\begin{figure}
\includegraphics[width=\columnwidth]{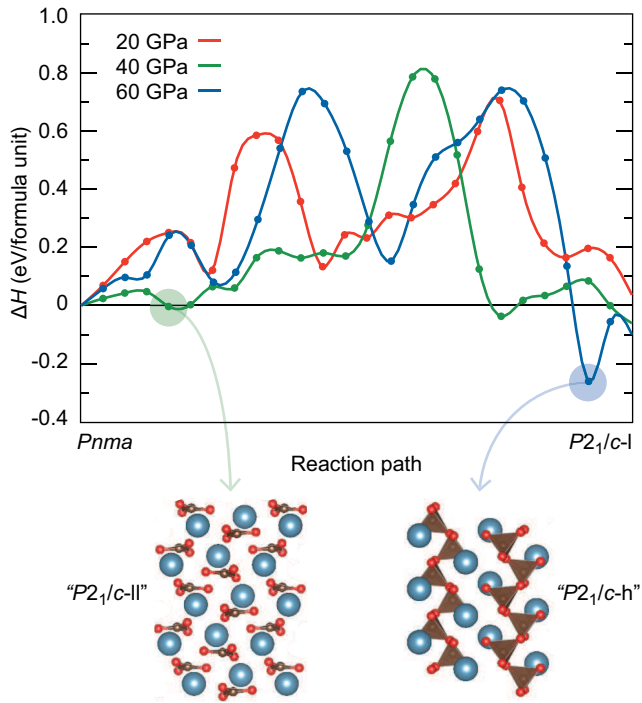}
\caption{\label{NEB} Mechanisms connecting the aragonite $Pnma$ phase to the new $P2_{1}/c$-l at 20 (red), 40 (green), and 60 (blue) GPa with enthalpies relative to the $Pmmn$ phase. Distances along the pathway are taken as the Euclidean norm between the images normalized so the defined phases lie at unit values. Appearance of $P2_{1}/c$-ll (green arrow) and $P2_{1}/c$-h (blue arrow) structures along mechanism pathways are highlighted.}
\end{figure}

The transformation between $Pnma$ to $P2_{1}/c$-l follows a similar trajectory at each pressure as can be seen in Fig. \ref{NEB}, and is marked by three processes: the rotation of the CO$_3^{2-}$ groups, a shifting of the Ca$^{2+}$ between atomic planes along the defined \textit{b} axis, and distortion of the unit cell to accommodate these motions. For reference, the $R\overline{3}c$ stacking planes are defined here as \textit{bc} planes perpendicular to the \textit{a} axis. The initial uphill steps from $Pnma$ have the system adopting a monoclinic cell angle. This provides an initial rotation of the CO$_3^{2-}$ groups commensurate with the monoclinic angle. After the first barrier, the CO$_3^{2-}$ then begin to rotate along a different axis and the cell angles return to nearly orthogonal. To accommodate this twisting of the CO$_3^{2-}$ groups the Ca$^{2+}$ switch stacking planes and migrate into a single \textit{ac} plane, until they become nearly planar at the minima halfway between $Pnma$ to $P2_{1}/c$-l. The next uphill trajectory has the Ca$^{2+}$ moving away from the single \textit{ac} plane while the box becomes more monoclinic, eventually reaching the global maximum along the pathway. We believe the roughness observed along this pathway is an artifact of the modest convergence criterion employed coupled with the use of variable spring constants between the images; tighter criterion should smooth out this climb. The global maximum has the largest unit cell volume of any of the structures, implying that the main enthalpic contribution is the $PV$ work the system must apply against the external pressure source to accommodate that larger volume. Following that highest enthalpy barrier, the cell volume drops below the final unit cell volume, but the higher enthalpies relative to the $P2_{1}/c$-l arise here from the repulsive sterics of non-optimally positioned atoms, and the cell axis perpendicular to the stacking plane elongates to allow for longer Ca$^{2+}$ distances. 

As this phase transition is kinetically hindered, it is worth estimating the transition temperature. If we assume that the potential energy barrier shown in Fig. \ref{NEB} is overcome entirely by thermally-induced kinetic energy ($\Delta{}H \sim{} E_{kin} = \frac{3}{2}Nk_{B}T$),\cite{Frenkel_Smit_2002} this estimates the transition temperatures to be 1091, 1216, and 1145 K for 20, 40, and 60 GPa, respectively. If a reasonable dimensionless rate constant ($\dot{N}/A$) of $1\times{}10^{-13}$ is assumed for this kinetically hindered solid-solid phase transition, the Arrhenius equation predicts nearly identical transition temperatures: 1094, 1219, 1148 K  for 20, 40, and 60 GPa, respectively. Altering the rate constant up or down by 2 orders of magnitude only affects these estimates by $\pm{}$ 100 K. These predicted barriers should be considered as upper bounds to the actual energetic barrier since this NEB connects only a single unit cell along the pathway defined in Fig. \ref{CO3angle}, and is by no means exhaustive of all possible transition pathways and other free energy concerns that may alter those values. With that being said, they do clearly demonstrate that while $P2_{1}/c$-l is the most enthalpically favorable structure at 40 GPa, there a clear necessity for the system to be driven at temperature to overcome the barrier into the P21/c-l phase − reinforcing the necessity of CO$_{2}$ laser annealing for investigating the phase progression in this and in other other geologically-relevant systems. A cold compression will not observe the $P2_{1}/c$-l phase, and that the phase has not yet been observed in over a decade of post-aragonite CaCO$_{3}$ experiments demonstrates that only an appropriately-designed experiment can unveil the true phase progression of mantle minerals.

Although the pathway in Fig. \ref{NEB} may not be the true minimum energy pathway, it does provide interesting physical insight about this complicated potential energy landscape pock-marked with the several similar, low-lying local minima defined in Fig. \ref{PH}. The first minimum about 20\%{} along the 40 GPa pathway very closely resembles the $P2_{1}/c$-ll phase predicted by AIRSS (shown in inset structure and highlighted by green circle in Fig. \ref{NEB}), and it having slightly higher enthalpy than $P2_{1}/c$-l here, unlike in AIRSS calculations (Fig. \ref{PH}), shows just how sensitive the magnitude of these small enthalpy changes are to the subtle differences in simulation set up. This minimum appears in all the structures, which seems obvious in retrospect as $P2_{1}/c$-ll can be best be described as a monoclinic distorted version of the $Pmna$ phase, where the rotations of the CO$_3^{2-}$ groups follow the monoclinic distortion of the cell. The barrier into this phase can be surmised as the energy penalty required to lower the symmetry of the cell from a favorable orthorhombic structure. Along the 60 GPa path, an intermediate structure actually becomes more enthalpically favorable than the $P2_{1}/c$-l phase. This structure corresponds to a lowered symmetry version of the $sp^{3}$ bonded $P2_{1}/c$-h phase (shown in inset structure and highlighted by blue circle in Fig. \ref{NEB}). At this point along the pathway, the counter-rotated CO$_3^{2-}$ groups become sufficiently close from the compressed cell volume that they react and bond to form the pyroxene-like chains of CO$_{4}$ $sp^{3}$ tetrahedra. The raise in enthalpy at the very end of the 60 GPa $Pnma$ to $P2_{1}/c$-l pathway comes from breaking a C-O bond to return to the $sp^{2}$ bound CO$_{3}^{2-}$ groups. Similar minima are observed along the 20 and 40 GPa pathways, however at those pressures the cell volume has not become sufficiently small to force the $sp^{3}$ bonding. The ability at higher pressures to form $P2_{1}/c$-h before forming $P2_{1}/c$-l short circuits the mechanistic pathway we have illustrated in Fig \ref{CO3angle}, allowing for a lower energy pathway connecting $P2_{1}/c$-h to stable higher pressure phases like $Pmmn$ akin to that studied in Ref. \onlinecite{Lobanov_2017}. Re-optimizations indicate that the NEB discovered $P2_{1}/c$-h structure becomes more enthalpically favorable by 50 GPa, precisely when the $P2_{1}/c$-l is no longer observed experimentally.

We also attempted to define similar pathways from both $Pmna$ and $P2_{1}/c$-l to $Pmmn$ using 2 unit cells, however in each case we were unable to obtain a satisfactorily converged pathway. One of the main reasons for this is the concerted rotation of the CO$_{3}^{2-}$ and re-arrangement of Ca$^{2+}$ groups necessitates that several groups come very close to one another creating large amounts of steric repulsion, and in some cases even leaving atoms directly on top of one another in the original images. This is likely because the matched interfaces of $Pmna$ and $P2_{1}/c$-l to $Pmmn$ are not ideal especially with the small cell volumes employed here, implying that larger scale mechanics such as diffusion would be required to achieve a phase transition into $Pmmn$. It should be noted that smallest energy barriers along the $P2_{1}/c$-l to $Pmmn$ pathway are double what was observed between $Pmna$ and $P2_{1}/c$-l, and at least double that for $Pmna$ to $Pmmn$. While these numbers are not converged and any inference is purely speculative, it does seem as though these transitions may require temperatures that are beyond those estimated to be in the mantle at 50-80 GPa (1250-1850 km, corresponding to 2054-2223 K) 
according to the PREM model.\cite{Dziewonski_1981} We intend to study the complicated $Pmmn$ phase transformation dynamics more thoroughly in a subsequent publication.

These complicated kinetics alongside our observation of experimental X-ray diffraction patterns which are not consistent with the reported $Pmmn$ phase casts some doubt over the claims made in recent years, \textit{i.e.} that aragonite will transform into a $Pmmn$ ``post-aragonite''. This doubt is further reinforced by the aforementioned lack of published structural refinements on the now-accepted $Pmmn$ structure. Furthermore, Ref. \onlinecite{Lobanov_2017} reveals a marked distinction in quality of diffraction data between their reported $Pmmn$ and $P2_{1}/c$-h samples, strongly evidenced by their ability to perform a Le Bail refinement on $P2_{1}/c$-h at 105 GPa, but reporting only a qualitative comparison between data at 83 GPa and a simulated $Pmmn$ pattern $-$ \textit{i.e.} experimental evidence for the existence of the $Pmmn$ structure is not as strong as evidence for each of the $P2_{1}/c$ structures.
The recent publication by Ref. \onlinecite{Gavryushkin_2017} reportedly features $Pmmn$ post-aragonite at 50 GPa alongside a proposed new structure with the $P2_{1}/c$-h unit cell, supplementing the argument that transformations into $Pmmn$ are severely kinetically hindered. 
With this in mind and with support from mechanistic calculations reported here showing the arrival of a structure above 50 GPa which is identical to $P2_{1}/c$-h, we postulate that $-$ in spite of its enthalpic favorability (Fig. \ref{PH}) $-$ $Pmmn$ may not be a real structure of CaCO$_{3}$ at mantle conditions. Tentative analysis of the post-$P2_{1}/c$-l data shown in Fig. \ref{stack} suggests that some mixture of the $P2_{1}/c$-l and -h may exist over some pressure range over which CaCO$_{3}$ is undergoing an isosymmetric transition.

\section{Conclusions}

We report an additional polymorph of CaCO$_{3}$ $-$ $P2_{1}/c$-l $-$ which is stable at pressures equivalent to a mantle depth around 1,000 km, first predicted to be stable by \textit{ab initio} random structure search \cite{Pickard_2015} and now realized by utilizing direct annealing of the mineral with 10.6 $\mu$m radiation from a CO$_{2}$ laser with \textit{in situ} X-ray diffraction. Above 50 GPa, the $P2_{1}/c$-l phase transforms into a structure which cannot be indexed as the formally accepted ``post-aragonite'', despite its apparent stability when analyzed with QHA calculations. Investigation of the reaction pathways between $P2_{1}/c$-l and $Pmmn$ ``post-aragonite'' sees a rising kinetic barrier with increasing pressure, slightly exceeding the temperatures estimated at equivalent depths inside the mantle by the PREM model,\cite{Dziewonski_1981} which may suggest an absence of $Pmmn$ CaCO$_{3}$ in the Earth's interior. Further work is required to formally investigate the phase progression of $P2_{1}/c$-l CaCO$_{3}$ upon further compression, but AIRSS and g-SSNEB calculations tentatively suggest that an isosymmetric transition from $P2_{1}/c$-l to $P2_{1}/c$-h may be able to facilitate direct $sp^{2}$-$sp^{3}$ conversion within mantle carbonates.

\begin{acknowledgments}
Data required to reproduce the calculation can be found at http://dx.doi.org/10.7488/ds/2289. This research was sponsored in part by the National Nuclear Security Administration under the Stewardship Science Academic Alliances program through DOE Cooperative Agreement \#DE-NA0001982. Portions of this work were performed at HPCAT (Sector 16), Advanced Photon Source (APS), Argonne National Laboratory. HPCAT operations are supported by DOE-NNSA under Award No. DE-NA0001974 and DOE-BES under Award No. DE-FG02-99ER45775, with partial instrumentation funding by NSF. APS is supported by DOE-BES, under Contract No. DE-AC02-06CH11357. C.J.P. acknowledges financial support from the Engineering and Physical Sciences Research Council (EPSRC) of the UK under Grant No. EP/P022596/1. C.J.P. is also supported by the Royal Society through a Royal Society Wolfson Research Merit Award.
\\

\end{acknowledgments}

\onecolumngrid

\begin{table}[t!]
\centering
\caption{\label{table}Crystallographic parameters of $P2_{1}/c$-l CaCO$_{3}$ from experiment and DFT.}
\begin{ruledtabular}
\begin{tabular}{ccccccccccc}
& Pressure &  & Lattice parameters && \multicolumn{3}{c}{Atomic coordinates} \\ &(GPa) & Space group & (\AA{}, deg.) & Species & x & y & z
\\
\hline\\
Experiment & 41.3 & $P2_ 1/c$ & $a=4.85656$ \hspace{0.3cm} $b=3.34107$ \hspace{0.3cm} $c=12.09737$ & Ca & 0.08584 & -0.19861 & 0.40372\\
& & &  $\alpha=90.0$ \hspace{0.3cm} $\beta=123.3300$ \hspace{0.3cm} $\gamma=90.0$ & C & -0.53927 & -0.05723 & 0.13886\\
& & & & O(1) & 0.27374 & -0.19485 & 0.17202\\
& & & & O(2) & -0.24458 & -0.18693 & 0.18103\\
& & & & O(3) & -0.66300 & -0.78832 & 0.04896\\ 
\\\\
Simulation&40& $P2_1/c$ & $a=4.71660$ \hspace{0.3cm} $b=3.34070$ \hspace{0.3cm} $c=12.42170$ & Ca & 0.09381 & -0.18682 & 0.39673\\
& & & $\alpha=90.0$ \hspace{0.3cm} $\beta=123.2500$ \hspace{0.3cm} $\gamma=90.0$ & C & -0.53927 & -0.05723 & 0.13886\\
& & & & O(1) & 0.27374 & -0.19485 & 0.17202\\
& & & & O(2) & -0.24458 & -0.18693 & 0.18103\\
& & & & O(3) & -0.66300 & -0.78832 & 0.04896\\ 
\end{tabular}
\end{ruledtabular}
\end{table}

\twocolumngrid


%

\end{document}